\documentclass[twocolumn,english,showpacs,aps,pra,floats,groupedaddress,superscriptaddress]{revtex4}
\usepackage[]{fontenc}
\usepackage[latin1]{inputenc}
\usepackage{graphicx,subfigure,dcolumn,bm}
\usepackage{amsfonts,amsmath,amssymb,epsfig,babel}
\newcommand{\bra}[1]{\langle #1 |}
\newcommand{\ket}[1]{| #1 \rangle}
\newcommand{\be}{\begin{equation}}
\newcommand{\ee}{\end{equation}}
\newcommand{\beq}{\begin{eqnarray}}
\newcommand{\eeq}{\end{eqnarray}}

\makeatother

\begin{document}

\title{Phase diagram of imbalanced strongly interacting fermions on a one dimensional optical lattice}

\author{Alberto Anfossi}
\affiliation{Dipartimento di Fisica dell'Universit\`a di Bologna, Viale Berti-Pichat 6/2, I-40127, Bologna, Italy}
\author{Luca Barbiero}
\affiliation{Dipartimento di Fisica del Politecnico, Corso Duca degli Abruzzi 24, I-10129, Torino, Italy}
\author{Arianna Montorsi}
\affiliation{Dipartimento di Fisica del Politecnico, Corso Duca degli Abruzzi 24, I-10129, Torino, Italy}


\date{20 May 2009}

\begin{abstract}
We show that the Hubbard Hamiltonian with particle-assisted tunneling rates --recently proposed to model a fermionic mixture near a broad Feshbach resonance-- displays a ground state phase diagram with superfluid, insulating, and phase separated regimes. In the latter case, when the populations are balanced the two phases coexist in microscopic antiferromagnetic domains. Macroscopic phase segregation into a high-density superfluid of molecules, and a low-density Fermi liquid of single atoms appears in the density profile above a critical polarization $p_c$.
\end{abstract}

\maketitle

\section{Introduction}
The recent experimental observation of normal-superfluid transition in two-component Fermi gases \cite{ZSSK,ZSSK2,SSSK} has attracted a great deal of attention on the subject of strongly interacting confined fermionic atoms. In this context, an important result consisted in the observation on the density profile of phase separation (PS) between a superfluid core and an external polarized shell, depending on the population imbalance $p$ and the interaction strength across a Feshbach resonance \cite{PAal,Shin}. The theoretical investigation of PS in the framework of uniform Fermi gases \cite{PIGI} provided an efficient description of the phase diagram along the BCS-BEC crossover. In particular, homogeneous phases of fully polarized and partly polarized normal gases or of unpolarized and partly polarized superfluids have been recognized, as well as the coexistence of some of these phases by varying the gas polarization and the interaction strength between atoms.

A description of the above phenomena when the fermionic atoms are confined by anisotropic optical lattices would represent a major achievement. On the theoretical side, it has been speculated that, with respect to real materials, these systems could provide a much neater realization of the Hubbard model \cite{JAZO,LEal}. The very recent experimental observation \cite{Jetal,Bloch} of the formation of an insulating phase in a repulsively interacting two-component Fermi gas on an optical lattice confirms this hypothesis, thus, enabling the use of such systems as a laboratory to investigate the onset of many phenomena in real materials, one for all high temperature superconductivity.

There is still no full agreement on the fact that Hubbard-like model Hamiltonians can support evidence of macroscopic phase segregation. Moreover, it has been observed that the Feshbach resonance, besides inducing the desired strong interaction among atoms \cite{Fesh}, can cause highly nontrivial effects \cite{Chin,STal,DIHO,duan1}, which should be effectively included in the lattice model. For instance, a broad resonance --typical of the experimental setups with fermionic atoms-- could generate a multiband distribution of the atomic population, which in turn affects the effective correlation between atoms on neighboring sites. In such a case, it was shown \cite{duan1} that a lattice resonance model is the generic model which would properly describe the system. The latter can be mapped \cite{duan2, WADU} into an extended Hubbard model with correlated hopping, previously known in literature as the Simon-Aligia model Hamiltonian \cite{SIAL}.

In this work, we provide exact analytical and numerical evidence that in the regime of strong interaction between fermions the Simon-Aligia model reproduces the scenario expected for ultracold fermionic gases, exhibiting normal, superfluid, insulating, and PS regimes. Based on such results, we also propose a simple explanation of the microscopic mechanism driving the transition to macroscopic phase segregation: while phase coexistence is present for appropriate parameters in a large range of filling values, the short range antiferromagnetic order masks this feature when the two populations are balanced and only for a critical polarization $p_c$ macroscopic phase segregation appears. Above it, a regime of breached pairs is entered.

\section{Lattice model Hamiltonian}

In the case of fermionic atoms with two internal states on a one-dimensional optical lattice, the Simon-Aligia model can be written as \cite{duan2}
\begin{widetext}
\be
    H_{SA} =  -\sum_{\sigma,\langle i,j \rangle} \left [t+\delta g(n_{i-\sigma} +n_{j-\sigma})+\delta t n_{i-\sigma}n_{j-\sigma}\right ] c_{i\sigma}^{\dagger}c_{j\sigma} + \frac{\Delta}{2}\sum_{i} n_{i}( n_{i}-1)\;,\label{ham_SA}
\ee
\end{widetext}
where $\sigma=\{\uparrow,\downarrow\}$ identifies the two internal states of the fermionic atoms ($-\sigma$ standing for the opposite of $\sigma$), $\langle i,j \rangle$ denotes two neighboring sites, and $n_i\doteq n_{i\uparrow}+ n_{i\downarrow}$. Here $t$  describes the direct hopping of atoms of a given population between neighboring  sites, while $\delta g=g-t$ and $\delta t=t+t_{ad}-2 g$ are the deviation from the direct hopping case (in which $\delta g=0=\delta t)$, induced by correlations in proximity of a wide Feshbach resonance. More precisely, $g$ describes the configuration tunneling, in which one atom is transferred onto a site already occupied by an atom of different specie: it amounts to coupling two fermionic atoms into a dressed molecule. $t_{ad}$ accounts for the motion of one atom between two already occupied sites: {\it de facto} it exchanges a dressed molecule and a fermionic atom located on neighboring sites. Finally, $\Delta$ is the energy cost of the dressed molecule, which works as an effective detuning parameter. Since in Eq. (\ref{ham_SA}) the number of atoms of both populations are conserved quantities, one can work at arbitrary fixed average number of atoms per site (filling) $n=N/L$ with $N= \langle\sum_i n_i\rangle$ ($0\leq n\leq 2$), and  population imbalance $p=\langle \sum_i n_{i\uparrow}-n_{i\downarrow}\rangle/N$ ($0\leq |p|\leq 1$).

In the context of highly-correlated fermionic materials, $H_{SA}$ has been widely studied \cite{vari,MON,ADM}. In particular, at $p=0$ it was recently shown \cite{MON} that --for the choice $\delta g=-t$ (i.e., $g=0$)-- its exact ground state phase diagram can be obtained, and it has been explored at $n=1$.  The structure of the latter is reminiscent for some aspect of the mentioned physics of cold fermionic gases, providing evidence of both PS and an insulating behavior for appropriate values of interaction parameters. Both features were confirmed by numerical analysis also for $0\leq g\leq -t_{ad}$ \cite{ADM}. Here we recast the exact solution of Ref. \onlinecite{MON} to the present context, allowing for arbitrary $p$, and we explore the role of population imbalance on the density profile at and away from the exact case.

\section{Results}

\subsection{$g=0$}

We first consider the $g=0$ case. A convenient representation of the model Hamiltonian (\ref{ham_SA}) is obtained by introducing  on-site Hubbard projection operators.  These are defined as $X^{\alpha\beta}_i\doteq\ket{\alpha}_i\bra{\beta}_i$, where $\ket{\alpha}_i$ are the states allowed at a given site $i$, and $\alpha=0,\uparrow,\downarrow,2$ ($\ket{2}\equiv \ket{\uparrow\downarrow}$). In terms of these operators the Hamiltonian  $H\doteq H_{SA} (\delta g=-t)$ reads as (up to constant terms)
\begin{eqnarray}
    H= \sum_{\langle i,j\rangle \sigma } \left [t X_i^{\sigma 0}X_j^{0\sigma}+ t_{ad}  X_i^{2\sigma}X_j^{\sigma 2}\right ] +  \Delta \sum_i X_i^{22}\; . \label{ham}
\end{eqnarray}
From a mathematical point of view, the choice $\delta g=-t$ implemented in Eq. (\ref{ham}) implies that also the total number of dressed molecules $N_d=\langle \sum_i X_i^{22}\rangle $ is a conserved quantity. Hence, the ground state of $H$ must be searched among the eigenstates of just its first two terms, namely, those with coefficients $t$ and $t_{ad}$. These terms are degenerate with respect to the population of the single atoms and their eigenstates do not depend on $p$.  One can show that in the thermodynamic limit, the ground state $\ket{\psi_{GS}(n)}$ consists of a high-density core of length $L_h$ with $N_d$ paired atoms and $L_h-N_d$ fermionic atoms, surrounded by a low-density shell of length $L-L_h$  occupied by $N-N_d-L_h$ single atoms, where $L_h$ and $N_d$ have to be determined variationally. Such a choice allows one to maximize the number of available low energy momenta. Explicitly,
\begin{equation}
\ket{\psi_{GS}(n)}= \ket{\psi_l (N-N_d-L_h)}_{L-L_h} \ket{\psi_h (L_h-N_d)}_{L_h} \; .
\end{equation}
Here $\ket{\psi_{\alpha} (N_f)}_{L}$ are the ground states of $N_f$ effective spinless fermions on a $L$-sites chain, moving in a background of empty sites $\alpha=l$ or dressed  molecules $\alpha=h$.

For $L_h=0=N_d$ the ground state coincides with $\ket{\psi_l (N_s)}_{L}$ and describes a normal (possibly partly polarized) Fermi liquid of single fermionic atoms, which we denote as $N_P$. Moreover, for $L_h=L$, the ground state becomes $\ket{\psi_h (L-N_d)}_{L}$, a (polarized) superfluid of dressed molecules and fermionic atoms ($SF_P$). In particular, for $L_h=N_d$, the latter consists just of dressed  molecules, and we denote the state as unpolarized superfluid ($SF_0$). Apart from these cases, $\ket{\psi_{GS}(n)}$ is always  characterized by PS. The two coexisting phases have different densities $n_l$, $n_h$, where $n_\alpha$ can be expressed in terms of the variational parameters $l_h\doteq L_h/L$ and $n_d\doteq N_d/L$ as
\be
    n_l=1-\frac{1-n+n_d}{1-l_h} \; ,\; n_h=1+\frac{n_d}{l_h}\; . \label{nlnh}
\ee
The boundaries of the PS regions in dependence of the physical parameters can be derived by investigating, at fixed $p$ and $n$, the dependence on $n_d$ and $l_h$ of the energy in the thermodynamic limit, namely, $ \lim_{L\rightarrow \infty} {1\over L} \bra{\psi_{GS}(n)}  H \ket{\psi_{GS}(n)}$. The ground state energy $e_{gs}(p,n)$ is obtained upon minimizing the latter with respect to $n_d$ and $l_h$
\begin{widetext}
\begin{equation}
    e_{gs}(p,n)= \min_{n_d,l_h} \left \{- {2\over \pi} \left[ (1-l_h)  \sin{\pi n_l} + l_h t_{ad} \sin{\pi (2-n_h)} \right]+\Delta n_d\right\}_p \; , \label{egs}
\end{equation}
\end{widetext}
where the expression in parenthesis is constrained by the actual value of the polarization $p$, since $n_d\leq n/2 (1-p)$. The optimal solution is obtained for $p=0$, in correspondence with $n_d=\bar n_d$ and $l_h=\bar l_h$. In order to discuss the scenario of imbalanced populations, we notice that the minimization equation for $n_d$ has formal solution $\bar n_d= {l_h\over \pi} \arccos {\frac{u-2\cos{\pi n_l}}{2 t_{ad}}}$.
While at $p=0$ this satisfies the constraint for $n_d$, in general it may miss such property for $p\neq 0$. In fact, this happens above a critical polarization
\be
    p_c=1-2 \bar n_d/n \; ,\label{pc}
\ee
when a regime of breached pairs is entered.
\begin{figure}[!h]
\includegraphics[width=90mm,keepaspectratio,clip]{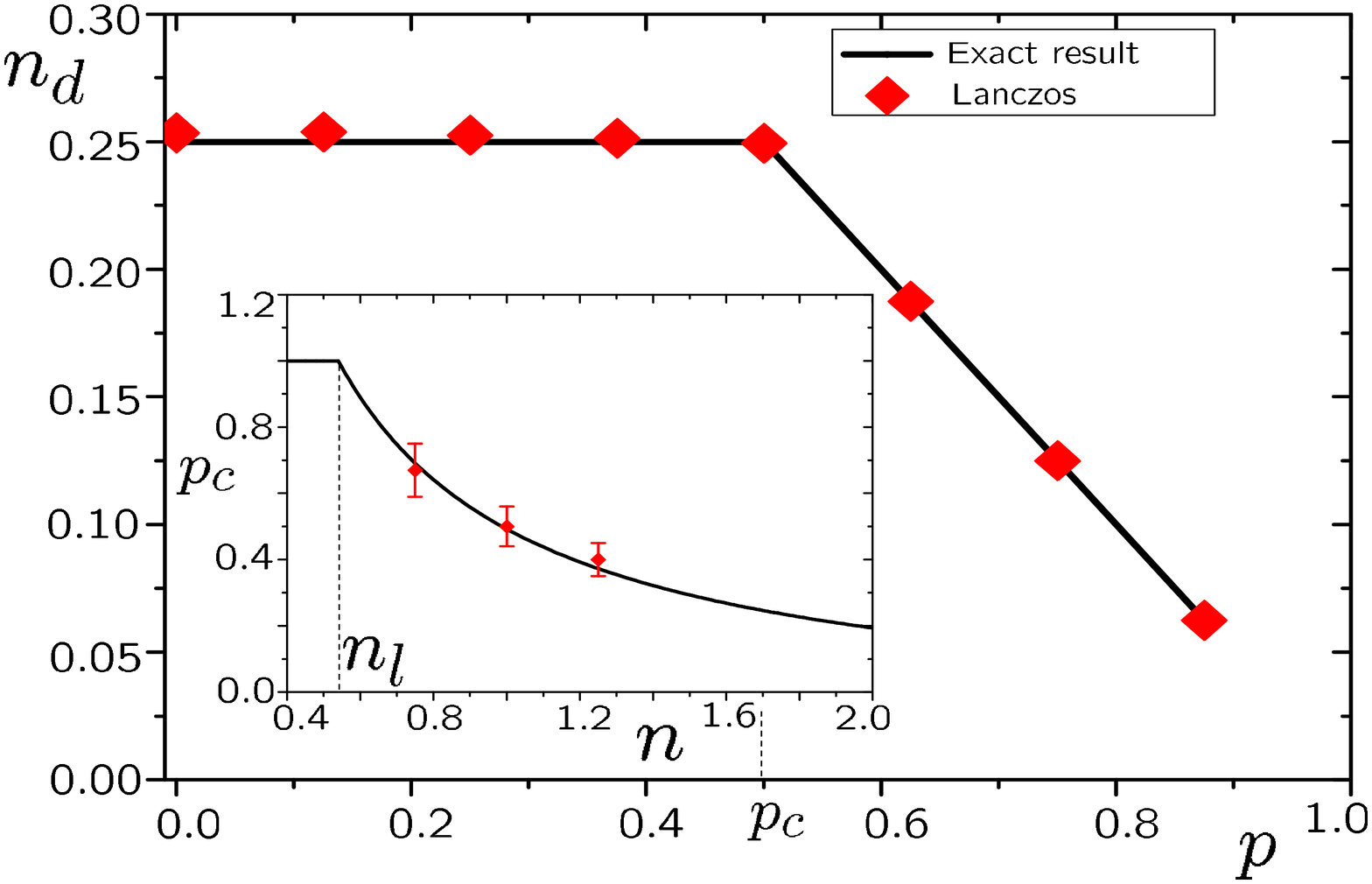}
\caption{(Color online) Number of pairs $n_d$ versus polarization $p$ at $\Delta=0$, and:  $\delta g=-t,\, \delta t=0.4t$ (continuous line, exact),  $\delta g=- 0.8t,\, \delta t=0$ (red diamonds, numerical). Inset: critical polarization $p_c$ vs filling $n$.}  \label{fig0}
\end{figure}
The system begins nucleating the normal phase within the superfluid to accommodate the excess polarization and, correspondingly, the number of dressed molecules $n_d$ diminishes with respect to the optimal value $\bar n_d$, $n_d=n/2 (1-p)$ for $p\geq p_c$, as shown in Fig. \ref{fig0}.

\begin{figure}[!h]
\includegraphics[width=90mm,keepaspectratio,clip]{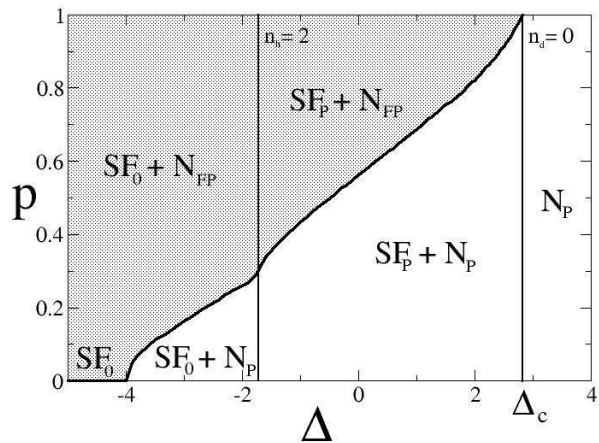}
\caption{Exact phase diagram in the $\Delta$-$p$ plane at $n=0.9$, $\delta g=-t$, and $\delta t= 0.4t$: the shaded region corresponds to macroscopic phase segregation in the density profile.}  \label{fig1}
\end{figure}

The ground-state phase diagram of the model described by $H$, as obtained from Eqs. (\ref{nlnh})-(\ref{pc}) is reported in Fig. \ref{fig1} in the $\Delta$-$p$ plane at a filling value $n<1$ and $t>t_{ad}> t/2$. Six different regions can be identified, depending on the values of $n_d$ and $l_h$ in the solution. In the regime of deep attractive detuning at $p=0$, the solution corresponds to the choice $\bar n_d=\bar l_h=n/2$: it describes an unpolarized superfluid of pairs, $SF_0$. In the same regime when $p\neq 0$, the solution becomes $l_h=n_d<\bar n_d$: the ground state nucleates in $SF_0$, a low-density normal phase of fully polarized Fermi liquid ($N_{FP}$). In this case, PS is favored with respect to the polarized superfluid, since  $t_{ad}\leq t$; for $t_{ad}>t$ the opposite scenario holds. When the detuning $\Delta$ is still attractive but moderate, as well as $p$, the solution reads $\bar n_d=\bar l_h< n/2$: the unpolarized superfluid coexists with a normal Fermi liquid ($N_P$). At lower-enough attractive $\Delta$ value, the solution becomes $\bar n_d<\bar l_h$. According to Eq. (\ref{nlnh}), this implies $n_h<2$: a regime is entered in which the superfluid is polarized ($SF_P$) since it contains $2-n_h$ unpaired atoms and still coexists with $N_P$. Such a region extends up to $\Delta_c$, which, in the present case, is moderately repulsive. Since the size $L-L_h$ of $N_P$ increases by enhancing $\Delta$, also the value of the critical polarization $p_c$ above which the normal phase becomes fully polarized increases accordingly. Above it, the $SF_P$ phase reduces its size by breaking some pair: this case corresponds again to $n_d<\bar n_d$. In both cases, the transition of the superfluid phase in the PS regime from $SF_0$ to $SF_P$ is identified by the line $n_h=2$. Moreover, the transition line to the breached pair regime (shaded region in Fig. \ref{fig1}) is characterized by $\bar n_d=n/2 (1-p)$. Finally, the uniform phase of normal Fermi liquid ($N_P$) is reached for $\Delta\geq \Delta_c$, the transition line being denoted by $\bar n_d=\bar l_h=0$.
The physical parameters $t_{ad}$ and $n$ also play a relevant role in the above scenario. The filling $n$ is crucial in the PS regimes, which is entered only for $n\geq n_l$. In particular, the $SF_P+N_P$ regime is the stable ground states in the range $n_l\leq n\leq n_h$, whereas the $SF_0+N_P$ state is stable up to $n_h=2$, $n_l$ and $n_h$ depending on the physical parameters through Eq. (\ref{nlnh}). Also the uniform phase changes by varying $n$: while for $n<1$ this is $N_P$, for $n=1$ it becomes insulating, and for $n>1$ it is $SF_P$. A thorough investigation shows that by decreasing $t_{ad}$, the phase diagram becomes more asymmetric with respect to the half filling configuration (not shown).

\subsection{$g\neq0$}

We now discuss the $g\neq0$ case. Since in this case no exact solution is available, one has to resort to numerical simulations for finding the ground state of $H_{SA}$. At  $p=0$, the situation was already investigated by means of the density-matrix renormalization-group algorithm \cite{ADM}: the essential features of the exact solution characterize a large range of $g$  values, namely, $0\leq g\lesssim - t_{ad} $, whereas a crossover region to the standard Hubbard regime is achieved when $g\geq -t_{ad}$ and still $t_{ad}<0$. The main difference with respect to the $g=0$ case is that the spin degrees of freedom become relevant. As a consequence, at $p=0$ when $0\leq g\lesssim - t_{ad}$ PS manifests itself at a microscopic level, by forming nanoscopic PS domains of size $\lambda$ (that turns out to be related to $n_d$) in which the coexisting phases rearrange their relative spins in order to maximize antiferromagnetic short-range order.

Given the above interpretative framework, we thus expect that $p\neq 0$ should not affect the results at $g\neq 0$ as far as the presence of phase coexistence is concerned. This is confirmed by the numerical data obtained by exact diagonalization on small clusters ($L=16$) for $n_d$ and $p_c$ at $g=0.2 t$, compared in Fig. \ref{fig0} with the exact results at $g=0$ and same $U$, $t_{ad}$, and $n$ values, showing accurate quantitative agreement. In both cases, increasing $p$ amounts to force the orientation of an increasing number of single atoms, while $n_d$ remains constant and equal to the $p=0$ optimal value up to $p= p_c$. Here, all the single atoms belong to one of the two internal states and, for $p>p_c$, some pair must break with respect to the optimal number $\bar n_d$.
\begin{figure}[!h]
\includegraphics[width=90mm,keepaspectratio,clip]{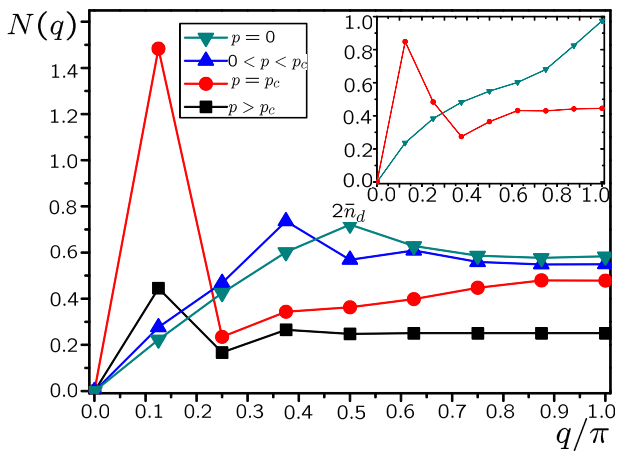}
\caption{(Color online) Static charge structure factor $N(q)$ at different $p$ values,   $L=N=16$, $\Delta=0$, $\delta t=0$, and $\delta g= - 0.8 t$ (figure), $\delta g =- 0.6 t$ (inset). In both cases, macroscopic phase separation emerges at $p\geq p_c$.}  \label{fig3}
\end{figure}
What is expected to change with respect to the $g=0$ case is instead the size of the PS domains. Since the number of single atoms available for such a short-range antiferromagnetic arrangement diminishes with $p$, $\lambda$ should increase correspondingly, so that for $p\geq p_c$ macroscopic phase segregation occurs also at $g\neq 0$. This is indeed confirmed by numerical investigation of the charge structure factor $N(q)\doteq \sum_r e^{i q r}  \left(\langle n_j n_{j+r}\rangle-\langle n_j\rangle\langle n_{j+r}\rangle \right)$ at half chain ($j=L/2$). The results are reported in Fig. \ref{fig3} at half filling. For $p\leq p_c$ the value at which $N(q)$ reaches its maximum moves smoothly from $ 2 \pi  n_d$ at $p=0$ to the lowest available $q$ value (i.e., $q=2\pi/L$) at $p=p_c$, where it remains also at higher $p$ values. In this case  macroscopic phase segregation is achieved. The inset of Fig. \ref{fig3} shows that the latter feature is maintained even in the crossover region $g\geq -t_{ad}> 0$, though in this case the behavior of $N(q)$ becomes closer to the standard Hubbard case for $p<p_c$.

\section{Conclusions}

In this paper, we provided exact analytical and numerical evidence that a lattice Hamiltonian recently proposed \cite{duan1,duan2} to mimic the physics of ultracold fermionic atoms on one-dimensional optical lattices close to a broad Feshbach resonance reproduces quite well the physics characteristic of these systems. In particular, aside to superfluid and insulating phases, a phase-separated regime is achieved in case of strongly interacting atoms. In this case, macroscopic phase segregation appears in the density profile above a critical polarization $p_c$ thanks to a peculiar mechanism. In fact, when PS is present in the system of balanced atoms, the coexisting phases form domains of microscopic dimension in order to implement antiferromagnetic short-range order. By increasing the imbalance of the population $p$, the size of the domains increases, to reach macroscopic phase segregation exactly at $p=p_c$ and above. The effect can be observed --for a large range of filling values around half filling-- also on the number of pairs $n_d$.

The model discussed here shows macroscopic phase segregation and Flude-Ferrell-Larkin-Ovchinnikov oscillations \cite{WADU}, both distinctive features in the investigation of systems of ultracold fermionic atoms. Therefore, our work gives evidence that Hubbard models with particle-assisted tunneling rates are more appropriate than the ordinary Hubbard model to describe the above physics when these systems are confined to anisotropic optical lattices. The present scenario should survive in higher dimension since its nature is to ascribe to the cooperative behavior of spin and charge degrees of freedom in the PS phase rather than to their separation.

\section{acknowledgments}
We thank F. Ortolani for the Lanczos code. This work was partially supported by the Grant No. PRIN 2005021773 of Italian MIUR.

\end{document}